# TRANSLATION

## ABOUT A FIRST ORDER TRANSFORMATION OF MAGNETITE AT 1 160 °C

par

### Claude CAREL*, Pierre VALLET [2]

(*Formerly Laboratoire de Chimie Générale A*)

\* University of Rennes1, 2 rue du Thabor CS 46510, 35065 RENNES Cedex, France : c-carel@orange.fr
[2] **(1906-1994)**   Sudoc instructions :  https://www.idref.fr/030362598#70

**Abstract**– In the literature on the wüstite/magnetite transformation, there are two distinct values of the enthalpy and entropy terms of the equilibrium between these two oxides. The formalism established previously for the calculation of thermodynamic properties of wüstite is taken as well as the adjustment of the boundary between the respective domains. A first-order transformation of the magnetite is demonstrated at 1 433 K or 1 160°C where the change in the reference enthalpy is $-9$ 990 J.mole$^{-1}$ at heating, with the corresponding entropy change being $-6.89$ J.K$^{-1}$.mole$^{-1}$. The approximate determination of enthalpy vs temperature is made for both varieties of magnetite. Two values of the molar heat content are deduced, with which a direct experimental value can be compared.

**Key words** – Iron oxides, high temperature, Magnetite, first order transition, ($\Delta$H°, $\Delta$S°, C$_p$°).

**RÉSUMÉ** – Dans la bibliographie relative à la transformation wüstite/magnétite, il apparaît deux valeurs distinctes des termes enthalpiques et entropiques de l'équilibre entre ces deux oxydes. Le formalisme établi précédemment pour le calcul des grandeurs thermodynamiques est repris de même que la rectification de la frontière entre les domaines respectifs. Une transformation du premier ordre de la magnétite est mise en évidence à 1 433 K ou 1 160 °C où la variation de l'enthalpie de référence vaut $-9$ 990 J.mole$^{-1}$ à l'échauffement, la variation d'entropie correspondante étant de $-6,89$ J.K$^{-1}$.mole$^{-1}$. La détermination approchée de la variation de l'enthalpie *vs* la température est faite pour les deux variétés de magnétite. Deux valeurs de la capacité calorifique molaire s'en déduisent, auxquelles une valeur expérimentale directe peut être comparée.

**Mots-clés** – Oxydes de fer, haute température, Magnétite, transition du premier ordre, ($\Delta$H°, $\Delta$S°, C$_p$°).

## INTRODUCTION

The boundary between iron monoxide or wüstite W written FeO$x$, and magnetite Fe3O4 cannot be described simply. A 4-terms equation between the decimal logarithm $l'_1$ of oxygen pressure p′$_1$, and temperature T(K) at equilibrium with these two compounds was previously given [1]. The usual thermodynamic equations make it possible to deduce the variations  $\Delta$H°$_1$  and   $\Delta$S°$_1$ for the chemical equilibrium studied previously [2]

$$\frac{6}{4-3x_1} \text{FeO}_{x_1}\text{ (s)} + \text{O}_2\text{ (g)} = \frac{2}{4-3x_1}\text{Fe3O4 (s)} \qquad (1)$$

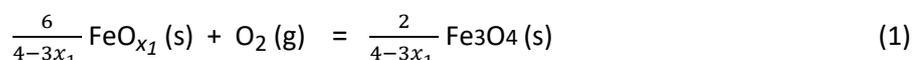

where the magnetite is assumed being stoichiometric along the boundary.

If $Y°$ symbolically represents either enthalpy or entropy, these reference properties can be calculated using the equation already proposed in [3], the index 1 of composition parameter $x$ being here only relative to the equilibrium W/ $Fe_3O_4$

$$Y°_{Fe_3O_4} = (2 - 1.5\, x_1)\, (\Delta Y°_1 + Y°_{O_2}) + 3\, Y°_{FeO_{x_1}} \qquad (2)$$

The calculation of enthalpy and entropy $Y°_{FeO_{x_1}}$ of the wüstite $FeO_{x_1}$ on the boundary with magnetite was also made explicitely in [3].

In the course of computing, the reference enthalpy and entropy of magnetite $Y°_{Fe_3O_4}$ show abnormal variations between 1 150 and 1 180°C. It appears also that the molar heat content under constant pressure of magnetite is noticeably smaller than that given by COUGHLIN, KING and BONNICKSON [4].

On the other hand, RIZZO, GORDON and CUTLER [5] in Table III, page 269, contrast their 900-1 200 °C results with those of the authors [6] for the 900-1 250 °C interval, and those of DARKEN and GURRY [7] above 1 100°C. However, the values of $l'_1$ obtained by RACCAH [8] are consistent with those of RIZZO et alii [4]. Those of one of us [9] confirmed the data of DARKEN and GURRY [7], which are excellent as indicated in [10].

## BASIC PRESENTATION

Let $(\log_{10} pO_2)_1$ abridged as $l'_1$ be the decimal logarithm of equilibrium oxygen pressure on the boundary W/$Fe_3O_4$. By combining 7 measurements of $l'_1$ considered to be the best ones performed above 1 200°C, the equation for the boundary between wüstite and magnetite is found as following

$$l'_1 = -(35\,336.45 \pm 184)\, T^{-1} + 14.844\,61 \qquad (3)$$

The linear correlation coefficient between $l'_1$ and $1/T$ is 0.999 993 and the confidence interval with which $l'_1$ is given by this equation at the usual probability threshold of 0.05 is $\pm 0.007\,9$.

Similarly, a second set of good measurements of $l'_1$ below 1 130 °C leads to the equation

$$l'_1 = -(33\,385.61 \pm 79)\, T^{-1} + 13.483\,31 \qquad (4)$$

The linear correlation coefficient between $l'_1$ and $1/T$ is 0.999 998 and the confidence interval with which $l'_1$ is given by this equation is $\pm 0.021\,3$.

The equations (3) and (4) are significantly different and give the same value $l'_1 = -9.813\,39$ for T = 1 433.1 K. As a guide, the equation of DARKEN and GURRY [7] recalled by RIZZO et alii (5) in the same Table III and their own have a common value of $l'_1$, equal to $-9.943$ for T = 1 425.5 K. These values are very close to those found here.

Equation (3) calculates the corresponding reference enthalpy variation $\Delta H°_1 = -676\,503$ J.mole$^{-1}$ and the entropy variation $\Delta S°_1 = -284.\,175$ J.K$^{-1}$.mole$^{-1}$ of reaction (1) above 1 433 K. Similarly, equation (4) provides $\Delta H°_2 = -639\,155$ J.mole$^{-1}$ and $\Delta S°_2 = -258.\,133$ J.K$^{-1}$.mole$^{-1}$ for the same reaction (1), but below 1 433 K.

Note that these variations of reaction enthalpy or entropy are multiplied by the factor $(4 - 3x_1)/2$ in equation (2) where they contribute to the calculation of reference enthalpy or entropy of magnetite. The numerical value of this factor is 0.267 493.

On the other hand, at 1 433 K at the common point of the two arcs of the boundary between wüstite and magnetite, the same variety of wüstite $W_2$ is in equilibrium with two varieties of magnetite, since on both sides of this temperature, in equation (2) will show either one or the other of these two values of enthalpy or entropy of the reaction. The reference enthalpy difference of the magnetite at 1 433 K will therefore be

$$H°_1 - H°_2 = (2 - 1.5x_1)\,(\Delta H°_1 - \Delta H°_2) \tag{5}$$

Similarly, the corresponding reference entropy difference will be

$$S°_1 - S°_2 = (2 - 1.5x_1)\,(\Delta S°_1 - \Delta S°_2) \tag{6}$$

The numerical calculation gives $H°_1 - H°_2 = -9\,990$ J.mole$^{-1}$ and $S°1 - S°2 = -6.971$ J.K$^1$.mole$^{-1}$. It is found that $-9\,990/1\,433 = -6.971$ J.K$^{-1}$.mole$^{-1}$. From a phenomenological point of view, this result characterizes a transformation of first order of the magnetite at 1 433 K or 1 160 °C. The transition from magnetite 2 (below 1 433 K) to magnetite 1 (above 1 433 K) is done with decreasing enthalpy and entropy. It is exothermic*.

* Erratum *published in **Bull. Soc. Sci. Bretagne,*** 54 (1-4), 1982 p. 113, indicating that the term «endothermic» should be replaced by the term «exothermic», is no longer applies.

## CONSEQUENCES

That is easily understood that reference enthalpy and entropy of magnetite can be calculated at various temperatures, on both sides of 1 160°C, using the above equations, some previous equations published before in [3], and numerical data in [1]. For oxygen and iron, we used thermodynamic data from the National Bureau of Standards (2nd edition, 1971).

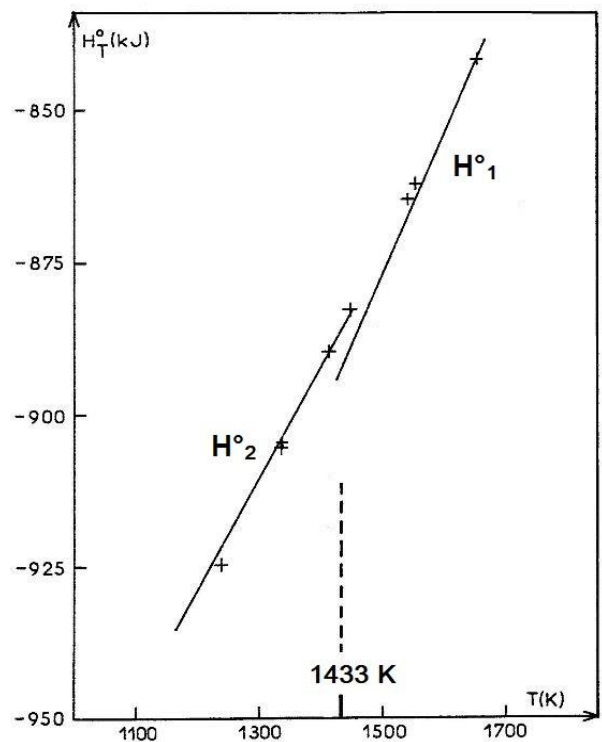

Graph resulting from equations (7) and (8) below.
Crosses are obtained from results of authors [4] when choosing $H°_2(298) = -112\,100$ J.mole$^{-1}$.

Above 1 160°C, for magnetite 1, we found in first approximation that $H°_1$ in J.mole$^{-1}$ is a linear function of T according to

$$H°_1 = (232.643 \pm 2.405)\,T - 1\,225\,687 \qquad (7)$$

The linear correlation coefficient between $H°_1$ and T is 0.999 933 and the confidence interval with which $H°_1$ is given by this equation is $\pm 460$ J.mole$^{-1}$.

Below 1 160°C for magnetite 2, as a first approximation $H°_2$ is also a linear function of T

$$H°_2 = (182.167 \pm 4.636)\,T - 1\,147\,430 \qquad (8)$$

The linear correlation coefficient between $H°_2$ and T is 0.999 430 and the confidence interval with which $H°_2$ is provided by equation (8) is $\pm 1\,192$ J.mole$^{-1}$.

In the above graphical representation of equations (7) and (8), two lines are obtained, of neighboring slopes and slightly shifted one in relation to the other. If the values of COUGHLIN *et alii* [4] are plotted on the figure, it can be seen that the representative points are placed in the vicinity of the two lines and that the single equation proposed by these authors is represented by an average line included in the angle of these lines. On the other hand, the molar heat content under constant pressure of 201 J.K$^{-1}$.mole$^{-1}$ given by these authors is intermediate between those immediately deduced from equations (7) and (8), namely 233 J.K$^{-1}$.mole$^{-1}$ for magnetite 1 and 182 J.K$^{-1}$.mole$^{-1}$ for magnetite 2.

This thermodynamic characterization of a magnetite transformation calls for the internal coherence in a carefully selected set of experimental results related to wüstite and on the wüstite/magnetite boundary. Anomalies in some series of results or disparities between them are thus found to get a satisfactory explanation.

Nevertheless, an improved version of this new disclosure of a first order transition in magnetite is envisaged in the next future, taking into account the nonstoichiometry of the oxide.

## ENDNOTE

- Additional references (1982, 1991)

In the literature, papers were found of which those by R. Dieckmann related to "Nonstoichiometry and Point Defect Structure of Magnetite ($Fe_{3-\delta}O_4$)" with the graph (T°C, [O/Fe]) of the boundary $Fe_{1-z_1}O/Fe_{3-\delta_1}O_4$. Initially Fig. 2 p. 94 in Materials Science Monographs, 10, Reactivity of Solids, Proceed. (Preprints) 9$^{th}$ I. S. R. S., Cracow, Sept. 1-6, 1980, Elsevier PWN Warszawa, I-10 p. 67-71, 1982, displays a simplified graph (T °C, [O]/[Fe]) of this boundary. Then, *the smoothed adapted graph* in Fig. 8 p. 118 *in* Ber Bunsenges Phys Chem, 86 (2), p. 112-118, 1982, (doi.org/10.1002/bbpc.19820860205) displays two variation regimes. Its graphically survey gives: I: between (1 400 °C, $\delta_1 = 1.3287$ and ~1 175 °C, $\delta_1 \approx 1.3313$). II: between (~ 1 175 °C, $\delta_1 \approx 1.3313$ and 848°C, $\delta_1 = 1.330$).

In a large and detailed Review, "Phase Diagram Evaluation, The Fe-O (Iron-Oxygen) System", J. of Phase Equil. 12 (N° 2) 1991, p. 169-200 (doi.org/10.1007/BF02645713), H. A. Wriedt makes p. 179 this comment concerning the boundary W/$Fe_{3-\delta_1}O_4$ by Dieckmann's paper: "the *increase* in deviation (Fe *excess*) of O concentration below the stoichiometric, a minimum deviation was depicted at about 1 200 °C, with temperature decreasing (*1 200 to 848 °C*) – what – *is unexpected*".
Now an opposite more sweeping (complex) situation at T > 1200 °C is observed in Fig. 8. Such a deviation could be attributed to an oxygen lattice deficit.

Would these two successive regimes of variation without a gap of $\delta$ in the curve at the transition correspond to distinct phases, magnetites 2 and 1 ? See further work in "Development" below by Vallet-Carel (1987).

- Development (1987)

P. Vallet, C. Carel, Propriétés thermodynamiques et transformations dans la magnétite non-stoechiométrique en équilibre avec les wüstites, *Revue de Chimie minérale*, t. 24 N°6, p. 719-737, 1987. To be translated into English.

**Citations**:

- Ref. [Ø] *is present in* (the bibliography of) *paper* "Kristina i. LiLova, CaroLyn i. PearCe, ChristoPher GorsKi, Kevin M. rosso, and aLexandra navrotsKy, Thermodynamics of the Magnetite-ulvöspinel (Fe₃O₄-Fe₂TiO₄) solid solution, American Mineralogist, 97 (8-9) p. 1330-8, 2012"
doi.org/10.2138/am.2012.4076

- Ref. [47] *in* (*the bibliography of*) *paper* "P. Vallet, C. Carel, Thermodynamic Properties of the Wüstites Wi and W'i from Thermogravimetric Data at Equilibrium, hal-03083695, v1  on January 28, 2021"

_______________________________